\documentclass[]{aa}

\usepackage{graphicx,epsf,psfig,epsfig}
\usepackage{natbib,multirow,amssymb}
\usepackage{txfonts}
\bibpunct{(}{)}{;}{a}{}{,}

\begin{document}
   \title{Simultaneous multi-wavelength observations of
   GRS\,1915+105\thanks{Based on observations with INTEGRAL, an ESA
   project with instruments and science data centre funded by ESA
   member states (especially the PI countries: Denmark, France,
   Germany, Italy, Switzerland, Spain), Czech Republic and Poland, and
   with the participation of Russia and the USA.}
}



   \author{Y. Fuchs\inst{1}, J. Rodriguez\inst{1,2}, I.F. Mirabel\inst{1,3}, S. Chaty\inst{1,4}, M. Rib\'o\inst{1}, V. Dhawan\inst{5}, P.~Goldoni\inst{1}, P. Sizun\inst{1}, G.G.~Pooley\inst{6}, A.A. Zdziarski\inst{7}, D.C. Hannikainen\inst{8}, P. Kretschmar\inst{9,2}, B. Cordier\inst{1}, N. Lund\inst{10}}
   \offprints{Y. Fuchs, \email{yfuchs@cea.fr}}


   \institute{Service d'Astrophysique (CNRS FRE 2591), CEA Saclay, 91191 Gif-sur-Yvette
 Cedex, France 
          \and  
Integral Science Data Center, Chemin d'Ecogia, 16, CH-1290 Versoix, Switzerland
	  \and 
Instituto de Astonom\'\i a y F\'\i sica del Espacio / CONICET, 
cc67, suc 28. 1428 Buenos Aires, Argentina
          \and 
Universit\'e Paris 7, F\'ed\'eration APC, 2 place Jussieu, 75005 Paris, France
          \and  
National Radio Astronomy Observatory, Socorro, NM 87801, USA
          \and 
Mullard Radio Astronomy Observatory, Cavendish Laboratory, Madingley Road, Cambridge CB3 0HE, UK
	  \and
N. Copernicus Astronomical Center, Bartycka 18, 00-716 Warsaw, Poland
          \and  
Observatory, PO Box 14, FIN-00014 University of Helsinki, Finland
          \and  
Max-Planck-Institut fuer Extraterrestrische Physik, Giessenbachstrasse, 85748 Garching, Germany
          \and  
Danish Space Research Institute, Juliane Maries Vej 30, DK-2100 Copenhagen Oe, Denmark
}
\authorrunning{Y. Fuchs et al.}

   \date{Received  ; accepted}

   \abstract{We present the result of multi-wavelength observations of
   the microquasar GRS\,1915+105 in a plateau state
   with a luminosity of $\sim7.5\times10^{38}$\,erg\,s$^{-1}$ ($\sim$\,40\%
   L$_\mathrm{Edd}$),
   conducted simultaneously with the INTEGRAL and RXTE satellites, 
   the ESO\thanks{Based on
   observations collected at the European Southern Observatory, Chile
   (ESO N\degr 071.D-0073).}/NTT, the Ryle Telescope, 
   the NRAO\thanks{The National Radio Astronomy Observatory is a facility of
the National Science Foundation operated under cooperative agreement by
Associated Universities, Inc.} 
   VLA and VLBA, in 2003 April 2--3. 
   For the first time were observed concurrently in GRS\,1915+105 
   all of the following properties:
   a strong steady optically thick radio emission
   corresponding to a powerful compact jet resolved with the VLBA, 
   bright near-IR emission, a strong QPO at 2.5\,Hz in the X-rays
   and a power law
   dominated spectrum without any cutoff in the 3-400\,keV range.
   }

   \maketitle

\section{Introduction}
	Microquasars are galactic X-ray binaries exhibiting
	relativistic jets \citep{mirabelrodriguez99}. The microquasar
	GRS\,1915+105 has been extensively observed since this source
	is known to be extremely variable at all wavelengths 
	(see \citealt{fuchs03} for a review).
	It hosts the most massive known
	stellar mass black hole of our Galaxy with $M=14\pm4M_{\odot}$ 
	\citep{greiner01Nat}.  
	It was the first galactic source to show superluminal ejections
	\citep{mirabelrodriguez94} in the radio domain,
	which has enabled to give an upper limit of 11.2$\pm$0.8\,kpc
	to the distance of the source \citep{fender99}. 
	In addition to these arcsecond scale ejections, 
	GRS\,1915+105 sometimes produces
	a compact jet which has been resolved at
	milli-arcsecond scales in radio \citep{dhawan00}, corresponding
	to a length of a few tens of AU.  The presence of such a compact
	and quasi-steady radio jet is now commonly inferred in 
	the low/hard X-ray state of
	several microquasars 
	but it is rarely resolved (see \citealt{fender03} 
	and references therein).


	\indent We present here the first multi-wavelength campaign on
	GRS\,1915+105 involving the recently launched INTErnational
	Gamma-Ray Astrophysics Laboratory (INTEGRAL, 3\,keV--10\,MeV).
	This campaign was conducted by the 
	MINE\footnote{See http://elbereth.obspm.fr/$\sim$fuchs/mine.html.}
	(\mbox{Multi-$\lambda$} INTEGRAL NEtwork) international collaboration
        aimed at performing multi-wavelength observations of
        galactic X-ray binaries 
        simultaneously with the INTEGRAL
	satellite.  In Sect.~2 we present an overview of our campaign,
	our results are shown in Sect.~3 and discussed in Sect.~4. 

\section{Overview of the multi-wavelength campaign}
	We conducted a multi-wavelength observation campaign of
	GRS\,1915+105 in March-April 2003.
	In this letter we focus only on April 2--3, when we obtained
	data covering the widest range of frequencies, with
	the largest number of involved instruments observing
	simultaneously with INTEGRAL (Fig.~\ref{figlightcurve}).
	These observations are ToO (Targets of
	Opportunity) triggered by the MINE collaboration under the
	INTEGRAL Guaranteed Time Programme (PI Mirabel) 
        and related programmes on the other
        instruments\footnote{Except for the RXTE observations which
        are part of a larger simultaneous INTEGRAL/RXTE campaign on
        GRS\,1915+105 (PIs Hannikainen, Rodriguez), for which the RXTE
        planning team kindly agreed to schedule an observation
        during the INTEGRAL ToO presented here.}.
%
	The detailed analysis of the whole campaign, i.e. March
	24--25, April 2--3 and April 17--18 2003
	(Fig.~\ref{figmonitor}) including further instruments,
	will be presented in forthcoming papers.  We thus present here
	an overview of the results of a (nearly) simultaneous campaign
	involving the Very Large Array (VLA), the Very Long Baseline
	Array (VLBA) and the Ryle Telescope (RT) in radio, the ESO New
	Technology Telescope (NTT)
	in IR, the Rossi X-ray Timing Explorer (RXTE) and INTEGRAL
	in X and $\gamma$-rays.

   \subsection{INTEGRAL} 
	\indent The data from the Integral Soft Gamma Ray Imager
	(ISGRI, 15\,keV--1\,MeV, \citealt{lebrun03}), were reduced with
	the Off-line Scientific Analysis software (OSA) v2.0 following
	\citet{goldwurm03}.
	The total elapsed time resulted in $\sim$101\,ks
        divided in 46 independent science windows (scw).
	For each of them we produced images in 2 energy bands, 20--40\,keV
	and 40--80\,keV, where GRS\,1915+105 was clearly detected and
	from which we could obtain its position and flux.
        The spectral extraction was performed independently for every scw
        in 128 channels linearly rebined in the 20--1000\,keV range.
	Given the low level of variability (less than 20\%, see
	Fig.~\ref{figlightcurve}), the resultant spectra were combined
	with \emph{mathpha}, and fitted between 20 and 400\,keV with
	\emph{xspec} v11.2 (also used for the SPI and RXTE spectral
	fits).  Note that we also produced spectra with the officially
	available 13--1000\,keV matrix. The results show no
	significant differences when fitting the spectra 
	in the 20--400\,keV range.



	\indent For the Spectrometer on board Integral (SPI, 20\,keV--8\,MeV,
	\citealt{vedrenne03}),  after correction of the
	misalignment, we extracted simultaneously the spectra of 6 
	sources seen by ISGRI in the 20--40\,keV band.
	We extracted the spectra of both the sources and
	background with the standard \emph{spiros} programme \citep{skinner03},
	using the latest derived response matrix \citep{sturner03}.
	The observation being made in dithering mode, we were able to
	resolve the background in each energy bin and each detector
	independently. 

	Using OSA v2.0, 
	JEM-X (Joint-European Monitor for X-rays, 3--35\,keV, \citealt{lund03})
	spectra were extracted \citep{westergaard03} for
	the 42 scw where the source was within 5\degr
	of the pointing direction and then combined for a total
	exposure of $\sim$88\,ks. The latest responses were used but
	the effective area corrected by a factor of 2 for the known
	losses due to dead anodes and the exclusion of the outermost
	areas of the detector.

   \subsection{RXTE}
	The RXTE observations were reduced following the standard
	procedure for bright sources, using the {\sc lheasoft} package
	v5.2. See e.g. \citet{rod03}, for the details of the
	Proportional Counter Array (PCA, 2--60\,keV) 
	and High Energy X-ray Timing Experiment (HEXTE, 20--200\,keV)
	data reduction followed  for the spectral extraction and
	analysis (note that we only extracted spectra from the 
	proportional counter unit 0 and 2 for the PCA). 
	We also extracted 
	light curves with $\sim$4\,ms resolution 
	from the PCA (see Fig.~\ref{figlightcurve}), and produced power
	spectra with \emph{powspec}~1.0. 

   \subsection{Infrared}
	The near infrared observations were performed with
	the 3.58\,m ESO/NTT 
	through a ToO programme (PI Chaty) dedicated to observations
	of high energy transient sources.
	 The telescope was equipped with the IR
	spectrograph and imaging camera SOFI.
	We took images in the broad band filters $J$ ($1.247\pm0.290\,\mu$m), 
	$H$ ($1.653\pm0.297\,\mu$m) and $K_{\mathrm s}$ ($2.162\pm0.275\,\mu$m)
	with a 4.9$'$$\times$4.9$'$ field of view
	(0.292$''$/pixel).
	Nine images of 60\,s integration time were acquired in each filter
	and co-added.
	We also observed in the $K_{\mathrm s}$-band with high temporal
	resolution (see Fig.~\ref{figlightcurve}), taking 450 exposures 
	of 2\,s each, by averaging 9$\times$2\,s frames, and randomly
	offsetting in distance and direction between each frame, during
	nearly 2.3\,hours.
	The images were reduced using the {\sc iraf}
	suite. Each image was corrected by a
	normalized dome-flat field, and sky-subtracted by a sky image
	(median filter combination 
	of 9 consecutive images).  The data were then analysed using 
	\emph{apphot}.
%
	Absolute photometry was performed using 2 standard stars: 
	HST P499-E and HST S808-C \citep{persson98}.
	The conditions were photometric for
	most of the observations, the seeing being typically 0.8$''$.

   \subsection{Radio}
\begin{figure}[!tbp]
\centering
\hspace*{-0.7cm}
\includegraphics[width=9cm]{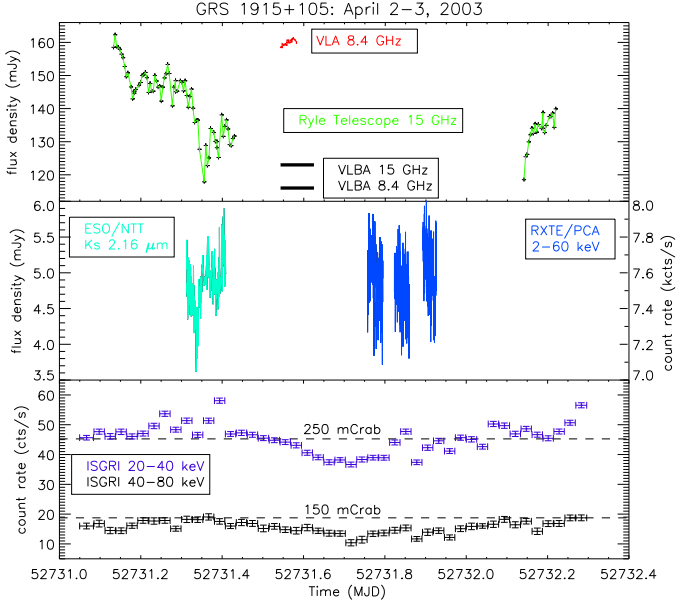}
\vspace*{-0.3cm}
\caption{Light curves of the multi-wavelength observations on April 2--3, 2003
involving INTEGRAL/ISGRI \& SPI (not plotted), RXTE/PCA \& HEXTE (not
plotted), ESO/NTT, RT, VLA and VLBA (total flux densities spanning the observing time).
}
\label{figlightcurve}
\end{figure}

We observed the source during two hours on April~~2, 2003 with the
VLBA at 2.0 and 3.6\,cm (ToO programme, PI Rib\'o). 
The observing strategy was the same as the one
described in \citet{dhawan00}. 
A detailed analysis
of all the VLBA data acquired during this campaign will be presented in
\citet{ribo03}.
The simultaneous radio light curve obtained with the  
VLA at 3.6\,cm (8.4\,GHz) during the first hour of the VLBA observations reveals a
nearly constant flux density of 160$\pm$1\,mJy (1$\sigma$ error, see Fig.~\ref{figlightcurve}), providing
good confidence in the obtained VLBA images.
%

\begin{figure}[!tbp]
\includegraphics[width=\columnwidth]{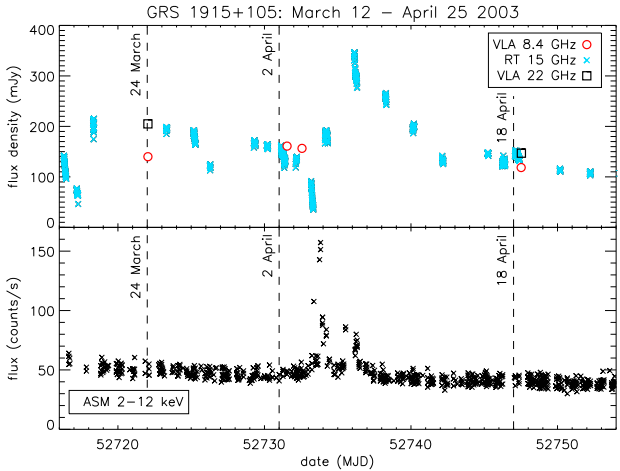}
\vspace*{-0.7cm}
\caption{Radio and X-ray flux monitoring of GRS\,1915+105 in March-April 2003. 
Radio: VLA at 8.4\,\&\,22\,GHz, RT at 15\,GHz. X-ray: quick-look
results provided by the ASM/RXTE team. The dashed lines indicate the
dates of our INTEGRAL and simultaneous multi-wavelength observations.
}
\label{figmonitor}
\end{figure}

     The Ryle Telescope observations at 15\,GHz are part of a
     continuing monitoring program (see \citealt{pooleyfender97} for
     more details).  The data are shown in
     Figs.~\ref{figlightcurve}~\&~\ref{figmonitor} with 5~min
     averaging; the typical uncertainty is 2\,mJy +\,3\% in flux
     scaling.


\begin{figure}[!tbp]
\includegraphics[width=\columnwidth]{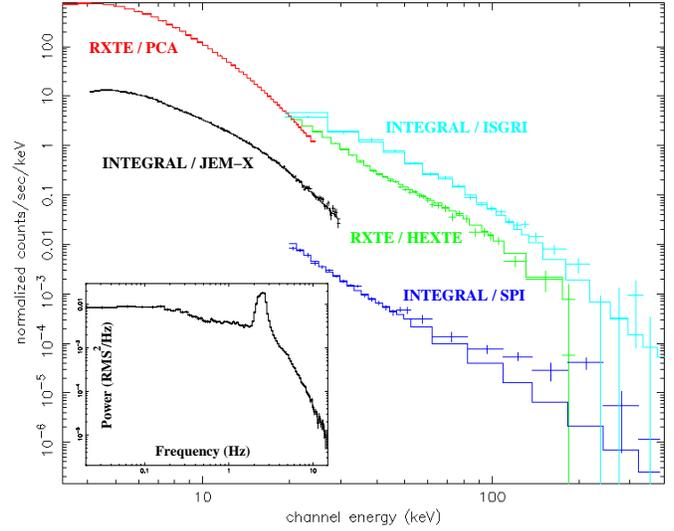}
\vspace*{-0.6cm}
   \caption{X-ray and $\gamma$-ray spectra of
   GRS\,1915+105 measured with RXTE (PCA and HEXTE) and INTEGRAL
   (JEM-X, ISGRI and SPI) on April 2, 2003. 
   Different sensitivities of the instruments lead to different levels
   of the spectra when plotted in count rates (which enables a better
   display).
   The structures at $E$$>$50\,keV in
   the SPI spectrum are instrumental background lines not adequately corrected.
   Continuous lines are the best fits with the models described in the text,
   showing consistent photon indexes among the different instruments.
   The PCA power density spectrum (inset)
   shows a clear QPO at 2.5\,Hz.}
   \label{figspectres}
\end{figure}
\begin{figure}[!tbp]
\includegraphics[width=\columnwidth]{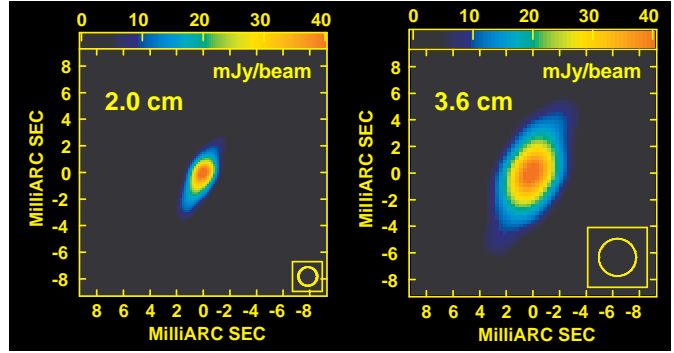}
\caption{
VLBA images at 2.0 \& 3.6\,cm  on 
April 2, 2003 showing the compact jet. Total integrated flux
densities are 123 \& 116\,mJy, respectively, 
in agreement with a slightly inverted spectrum. The convolving
beams
are 1.4 \& 2.8\,mas, respectively. 
1\,mas corresponds to 12\,AU at 12\,kpc distance. The rms noise in
both maps is 0.15\,mJy\,beam$^{-1}$.  
}
\label{figjet}
\end{figure}
\begin{figure}[!tbp]
\includegraphics[width=\columnwidth]{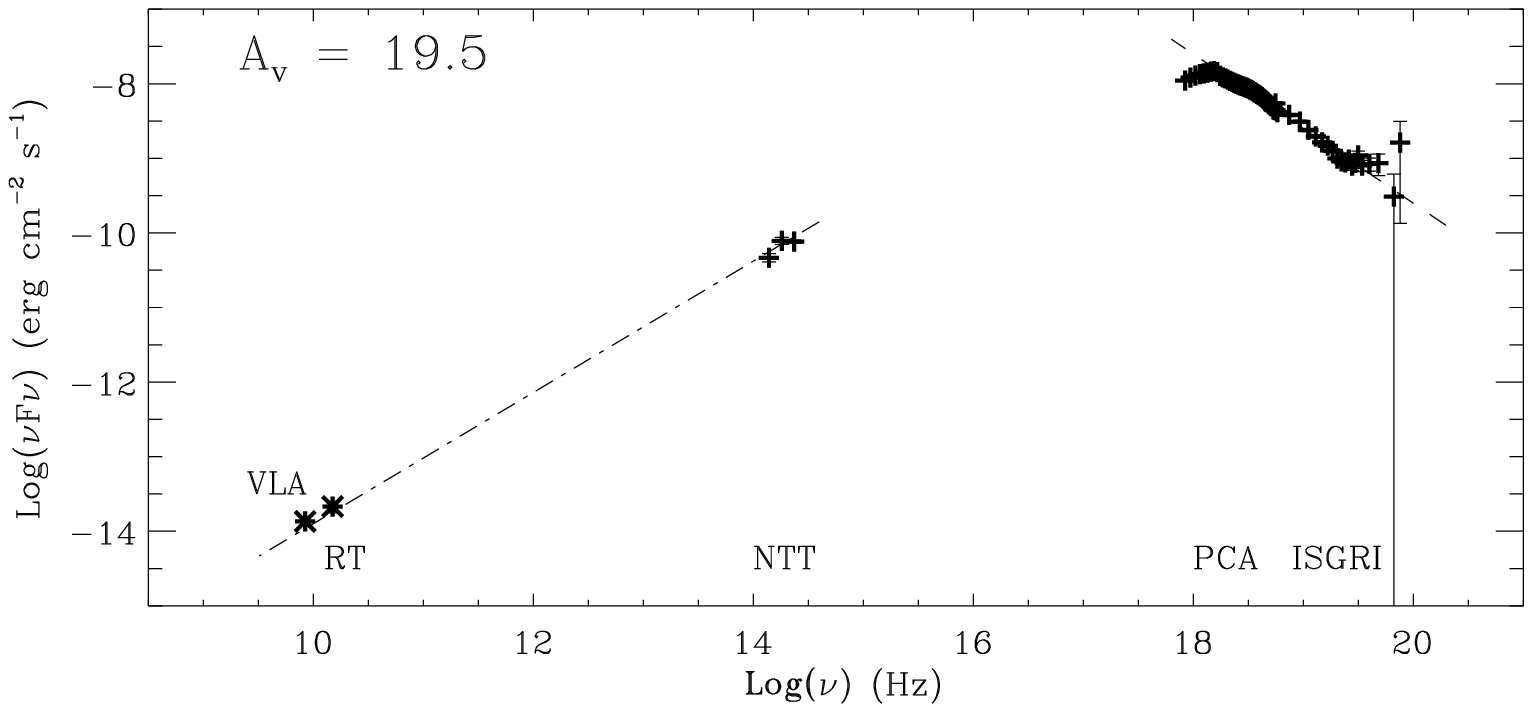}
\vspace*{-0.7cm}
\caption{
Spectral energy distribution of GRS\,1915+105 on April~2. The near-IR flux densities were dereddened with $A_V$\,=\,19.5 and using eq.\,1 of \citet{cardelli89}. The dot-dashed line illustrates the optically thick synchrotron emission from the jet as a power law with $F_\nu\propto\nu^{-0.12}$. The dashed line illustrates a power law with a photon index $\Gamma$=3. Error bars not shown are smaller than the symbol size.
}
\label{figsed}
\end{figure}

\section{Results}
	Figure~\ref{figmonitor} shows the general radio and X-ray behaviour 
	of GRS\,1915+105 during our observing campaign. 
	Despite X-ray and giant 
	radio flaring episodes that occurred in March--April 2003, our
	multi-wavelength observations took place during quasi-quiet
	periods, with a slowly decaying ASM flux
	$\sim50$\,cts/s and an unusually high radio level 
	($>$100\,mJy).

	The accumulated spectra obtained with INTEGRAL (ISGRI and SPI)
	and with RXTE (PCA and HEXTE) on April 2--3 are shown in
	Fig.~\ref{figspectres}.
	The ISGRI and SPI spectra are well jointly fitted in the 20--400\,keV
	range by a power law with a photon index of $\Gamma = 3.16\pm0.04$
	which is flatter than the $\Gamma=3.58\pm0.02$ index of the
	HEXTE spectrum in the 20--200\,keV range. This difference may
	be due to either instrument, since the background noise of
	HEXTE is uncertain when pointing in the galactic plane, and
	the calibration (e.g. background correction) of ISGRI and SPI
	are still in progress. Moreover, the HEXTE spectrum is built
	only on $\sim$6\,ks of observation whereas the INTEGRAL spectra are
	averaged over 101\,ks.


	The PCA + HEXTE spectrum (3--200\,keV, cf
	Fig.~\ref{figspectres}) was fitted by the sum of a multicolor
	blackbody model (diskbb), a power law and an iron line at $\sim$6\,keV
	assuming a interstellar absorption with
	$N_{\mathrm H}=5\times10^{22}$\,cm$^{-2}$. The fitted power law,
	with $\Gamma=2.94\pm0.01$, accounts for 77\% 
	of the total unabsorbed 3--20\,keV flux, 
	and is flatter than at higher energies where the soft 
	component has no longer influence. 
	The resulting inner disk temperature 
	is very high ($kT>3$\,keV) as already noticed by
	\citet{muno99,muno01} who obtained similar results while
	fitting the plateau state of GRS\,1915+105 (cf discussion),
	which may indicate either that the diskbb model is not valid
	here, or that other emission processes, such as Comptonization
	in thermal plasma \citep{zdziarski01}, are
	responsible for the low energy X-ray component.
	An analysis of the averaged spectrum obtained with JEM-X 
	gives spectral
	fits (4--30\,keV) consistent with the PCA.
	The estimated luminosity is
	$\sim7.5\times10^{38}$\,erg\,s$^{-1}$ corresponding to
	$\sim$40\% of the Eddington luminosity for a $14M_{\odot}$
	black hole. 
	As shown in Fig.~\ref{figspectres}, a very
	clear Quasi-Periodic Oscillation (QPO) at 2.5\,Hz with a 14\%
	rms level was observed in the RXTE/PCA signal.

	The VLBA high resolution images (Fig.~\ref{figjet}) show the
	presence of a compact radio jet with a $\sim$7--14\,mas length
	(85--170\,AU at 12\,kpc). This jet is very similar to the one
	observed by \citet{dhawan00} and is responsible for the high
	radio levels measured with the RT and the VLA
	(Fig.~\ref{figmonitor}) by its optically thick synchrotron
	emission.

	The source was fairly bright in near-IR 
	(compare e.g. to \citealt{chaty96})
	with apparent magnitudes 
	$J=17.19\pm0.06$, $H=14.46\pm0.12$ and $K_{\mathrm s}=12.95\pm0.14$.
	This corresponds to an excess of 75\% to 85\%
	in the $K_{\mathrm s}$-band flux compared to the $K$=14.5-15\,mag.
	of the K-M giant donor star of the X-ray binary \citep{greiner01}.
	The spectral energy distribution of Fig.~\ref{figsed} is
	compatible with a strong contribution to the near-IR bands from
	the synchrotron emission of the jet extending from the radio
	up to the near-IR. 
	However, the $J$, $H$, and $K_{\mathrm s}$ dereddened flux
	densities are not compatible with such a single power law
	emission, since
	a significant change in the slope appears
	at the $H$-band  by  dereddening with a 
	visible absorption $A_V=19.5$ 
	\citep{chapuis03}.  We point out here that while the value of
	the visible absorption is still a matter in the debate (see
	e.g. \citealt{fuchs03} and \citealt{chaty96}), this H band
	excess might be due to
	the sum of the
	different components adding to the jet emission in the
	near-IR, such as the donor-star, the external part of the
	accretion disc or a free-free emission.
	
	The detailed light curves of our observations
	(Fig.~\ref{figlightcurve}) show nearly quiet flux densities
	when compared to flares or oscillations, with variations
	$\lesssim$20\% at all wavelengths.
	The most interesting phenomenon is a moderate $\sim$25\%
	decrease (from 4.9 to 3.6\,mJy) in the $K_{\mathrm s}$ flux density 
	lasting 20\,min which precedes by 31\,min a $\sim$20\% decrease in
	the RT signal (from 145 to 118\,mJy) lasting 48\,min.  This 
	may be due to instabilities in the jet
	inducing an
	immediate synchrotron response in the IR and the delay being
	due to the time for the material along the jet to become
	optically thin to the radio emission \citep{mirabel98}. 

\section{Discussion}

	Bright radio emission ($F_\nu$\,$\gtrsim$\,100\,mJy at 15\,GHz)
	accompanied by steady X-ray emission (ASM $\sim$50\,cts/s)
	similar to what happened around April 2, 2003 were observed on
	several past occasions in GRS\,1915+105 
	(see e.g. Fig.\,1 of \citealt{muno01}).
	This state is known as the plateau
	state (\citealt{fender99,kleinwolt02} and references therein) 
	but is also called the radio loud low/hard X-ray
	state \citep{muno01} and type II state
	\citep{trudolyubov01}. It also corresponds to the $\chi_1$
	and $\chi_3$ X-ray classes of \citet{belloni00} 
	who used PCA color-color diagram, and indeed our April 2 
	observation appears as the $\chi_1$ class when plotted 
	in such a diagram.
	\citet{dhawan00} observed GRS\,1915+105 with the VLBA during
	the 1998 plateau state, and they found a compact radio jet
	very similar to the one shown in Fig.~\ref{figjet}.

	The strong QPO at 2.5\,Hz observed on April 2 is 
	consistent with the low/hard state of GRS\,1915+105.
	The presence of the 0.5--10\,Hz
	QPO seems correlated to a hard tail in the energy spectra 
	of the source \citep{markwardt99,muno99}, 
	although
	the energy dependence of the QPO amplitude might show a
	cut-off at high energy \citep{rod02}, indicating that the QPO
	is probably not related to a global oscillation of that hard
	component.

	The high energy emission of GRS\,1915+105 on April~2 is also
	consistent with the low/hard state of the source, with a power
	law dominated spectrum (77\% at 3--20\,keV), although always
	softer ($\Gamma$$\sim$3) than for the other BH binaries
	\citep{mcclintock03}. The INTEGRAL observations show
	that this power law spectrum extends up to 400\,keV without
	any cutoff during this plateau state, consistent with the
	observations with OSSE \citep{zdziarski01}.

	Here for the first time, we observed simultaneously all the
	properties of the plateau state of GRS\,1915+105 that were
	previously observed individually.
	We thus confirm the presence of a powerful compact
	radio jet, responsible for the strong steady radio emission and
	probably for a significant part of the bright near-IR
	emission, as well as a QPO (2.5\,Hz) in the X-rays 
	and a power law dominated X-ray spectrum with
	a $\Gamma$$\sim$3 photon index up to at least
	400\,keV. 
	Detailed fits of the RXTE and INTEGRAL spectra of GRS\,1915+105
	in this plateau state, to determine for example
	whether this power law is due to an inverse Compton
	scattering of soft disc photons 
	on the base of the compact jet 
	(see e.g. \citealt{fender99,raugreiner03})
	or not, will be studied in forthcoming papers.
	In our multi-wavelength March-April campaign, the source was
	observed essentially in the plateau state. In order to better
	understand the unusual behaviour of GRS\,1915+105, we need to
	carry out similar simultaneous broad-band campaigns during the
	other states, in particular during the sudden changes in the
	X-ray state that correspond to powerful relativistic
	ejection events.

	Detailed analysis and interpretation of all of our
	observations and their scientific implications will be
	presented, separately, in future articles.

\begin{acknowledgements}
Y.F. and J.R. acknowledge financial support from the CNES.
M.R. acknowledges support from a Marie Curie individual fellowship under
contract No.\,HPMF-CT-2002-02053.
D.C.H. acknowledges the Academy of Finland for financial support.
A.A.Z. has been supported by KBN grants 5P03D00821, 2P03C00619p1,2 and 
PBZ-054/P03/2001.
S.C. is grateful to skills and availability of the ESO staff for performing
ToO programmes, and particularly to the support astronomer Emanuela Pompei.
\end{acknowledgements}
\bibliographystyle{aa}
\bibliography{ref-1915-mine}

\begin{thebibliography}{33}
\expandafter\ifx\csname natexlab\endcsname\relax\def\natexlab#1{#1}\fi

\bibitem[{{Belloni} {et~al.}(2000){Belloni}, {Klein-Wolt}, {M{\' e}ndez}, {van
  der Klis}, \& {van Paradijs}}]{belloni00}
{Belloni}, T., {Klein-Wolt}, M., {M{\' e}ndez}, M., {van der Klis}, M., \& {van
  Paradijs}, J. 2000, \aap, 355, 271

\bibitem[{{Cardelli} {et~al.}(1989){Cardelli}, {Clayton}, \&
  {Mathis}}]{cardelli89}
{Cardelli}, J.~A., {Clayton}, G.~C., \& {Mathis}, J.~S. 1989, \apj, 345, 245

\bibitem[{{Chapuis} \& {Corbel}(2003)}]{chapuis03}
{Chapuis}, C. \& {Corbel}, S. 2003, \aap \ subm.

\bibitem[{{Chaty} {et~al.}(1996){Chaty}, {Mirabel}, {Duc}, {Wink}, \& {Rodr\'\i
  guez}}]{chaty96}
{Chaty}, S., {Mirabel}, I.~F., {Duc}, P.~A., {Wink}, J.~E., \& {Rodr\'\i guez},
  L.~F. 1996, \aap, 310, 825

\bibitem[{{Dhawan} {et~al.}(2000){Dhawan}, {Mirabel}, \& {Rodr\'\i
  guez}}]{dhawan00}
{Dhawan}, V., {Mirabel}, I.~F., \& {Rodr\'\i guez}, L.~F. 2000, \apj, 543, 373

\bibitem[{{Fender}(2003)}]{fender03}
{Fender}, R. 2003, in 'Compact Stellar X-Ray Sources', eds. W.H.G. Lewin and M.
  van der Klis, CUP, astro-ph/0303339

\bibitem[{{Fender} {et~al.}(1999){Fender}, {Garrington}, {McKay}, {Muxlow},
  {Pooley}, {Spencer}, {Stirling}, \& {Waltman}}]{fender99}
{Fender}, R.~P., {Garrington}, S.~T., {McKay}, D.~J., {et~al.} 1999, \mnras,
  304, 865

\bibitem[{{Fuchs} {et~al.}(2003){Fuchs}, {Mirabel}, \& {Claret}}]{fuchs03}
{Fuchs}, Y., {Mirabel}, I.~F., \& {Claret}, A. 2003, \aap, 404, 1011

\bibitem[{{Goldwurm} {et~al.}(2003){Goldwurm}, {David}, {Foschini}, {Gros},
  {Laurent}, {Sauvageon}, {Bird}, {Lerusse}, \& {Produit}}]{goldwurm03}
{Goldwurm}, A., {David}, P., {Foschini}, L., {et~al.} 2003, \aap\,subm.

\bibitem[{{Greiner} {et~al.}(2001{\natexlab{a}}){Greiner}, {Cuby}, \&
  {McCaughrean}}]{greiner01Nat}
{Greiner}, J., {Cuby}, J.~G., \& {McCaughrean}, M.~J. 2001{\natexlab{a}}, \nat,
  414, 522

\bibitem[{{Greiner} {et~al.}(2001{\natexlab{b}}){Greiner}, {Cuby},
  {McCaughrean}, {Castro-Tirado}, \& {Mennickent}}]{greiner01}
{Greiner}, J., {Cuby}, J.~G., {McCaughrean}, M.~J., {Castro-Tirado}, A.~J., \&
  {Mennickent}, R.~E. 2001{\natexlab{b}}, \aap, 373, L37

\bibitem[{{Klein-Wolt} {et~al.}(2002){Klein-Wolt}, {Fender}, {Pooley},
  {Belloni}, {Migliari}, {Morgan}, \& {van der Klis}}]{kleinwolt02}
{Klein-Wolt}, M., {Fender}, R.~P., {Pooley}, G.~G., {et~al.} 2002, \mnras, 331,
  745

\bibitem[{{Lebrun} {et~al.}(2003){Lebrun}, {Leray}, {Lavocat}, {Cr\'etolle},
  {Arqu\`es}, \& {Blondel}}]{lebrun03}
{Lebrun}, F., {Leray}, J.-P., {Lavocat}, P., {et~al.} 2003, \aap\,subm.

\bibitem[{{Lund} {et~al.}(2003){Lund}, {Brandt}, {Budtz-Jorgensen}, {Unknown},
  {Smith}, \& {Doe}}]{lund03}
{Lund}, N., {Brandt}, S., {Budtz-Jorgensen}, C., {et~al.} 2003, \aap\,subm.

\bibitem[{{Markwardt} {et~al.}(1999){Markwardt}, {Swank}, \&
  {Taam}}]{markwardt99}
{Markwardt}, C.~B., {Swank}, J.~H., \& {Taam}, R.~E. 1999, \apjl, 513, L37

\bibitem[{{McClintock} \& {Remillard}(2003)}]{mcclintock03}
{McClintock}, J.~E. \& {Remillard}, R.~A. 2003, in 'Compact Stellar X-Ray
  Sources', eds. W.H.G. Lewin and M. van der Klis, CUP, astro-ph/0306213

\bibitem[{{Mirabel} {et~al.}(1998){Mirabel}, {Dhawan}, {Chaty}, {Rodr\'\i
  guez}, {Marti}, {Robinson}, {Swank}, \& {Geballe}}]{mirabel98}
{Mirabel}, I.~F., {Dhawan}, V., {Chaty}, S., {et~al.} 1998, \aap, 330, L9

\bibitem[{{Mirabel} \& {Rodr\'\i guez}(1994)}]{mirabelrodriguez94}
{Mirabel}, I.~F. \& {Rodr\'\i guez}, L.~F. 1994, \nat, 371, 46

\bibitem[{{Mirabel} \& {Rodr{\'{\i}}guez}(1999)}]{mirabelrodriguez99}
{Mirabel}, I.~F. \& {Rodr{\'{\i}}guez}, L.~F. 1999, \araa, 37, 409

\bibitem[{{Muno} {et~al.}(1999){Muno}, {Morgan}, \& {Remillard}}]{muno99}
{Muno}, M.~P., {Morgan}, E.~H., \& {Remillard}, R.~A. 1999, \apj, 527, 321

\bibitem[{{Muno} {et~al.}(2001){Muno}, {Remillard}, {Morgan}, {Waltman},
  {Dhawan}, {Hjellming}, \& {Pooley}}]{muno01}
{Muno}, M.~P., {Remillard}, R.~A., {Morgan}, E.~H., {et~al.} 2001, \apj, 556,
  515

\bibitem[{{Persson} {et~al.}(1998){Persson}, {Murphy}, {Krzeminski}, {Roth}, \&
  {Rieke}}]{persson98}
{Persson}, S.~E., {Murphy}, D.~C., {Krzeminski}, W., {Roth}, M., \& {Rieke},
  M.~J. 1998, \aj, 116, 2475

\bibitem[{{Pooley} \& {Fender}(1997)}]{pooleyfender97}
{Pooley}, G.~G. \& {Fender}, R.~P. 1997, \mnras, 292, 925

\bibitem[{{Rau} \& {Greiner}(2003)}]{raugreiner03}
{Rau}, A. \& {Greiner}, J. 2003, \aap, 397, 711

\bibitem[{{Rib\'o} {et~al.}(2003){Rib\'o}, {Dhawan}, \& {Mirabel}}]{ribo03}
{Rib\'o}, M., {Dhawan}, V., \& {Mirabel}, I.~F. 2003, in prep.

\bibitem[{{Rodriguez} {et~al.}(2003){Rodriguez}, {Corbel}, \&
  {Tomsick}}]{rod03}
{Rodriguez}, J., {Corbel}, S., \& {Tomsick}, J. 2003, \apj \ in press

\bibitem[{{Rodriguez} {et~al.}(2002){Rodriguez}, {Durouchoux}, {Mirabel},
  {Ueda}, {Tagger}, \& {Yamaoka}}]{rod02}
{Rodriguez}, J., {Durouchoux}, P., {Mirabel}, I.~F., {et~al.} 2002, \aap, 386,
  271

\bibitem[{{Skinner} \& {Connell}(2003)}]{skinner03}
{Skinner}, G. \& {Connell}, P. 2003, \aap \ subm.

\bibitem[{{Sturner} {et~al.}(2003){Sturner}, {Shrader}, {Weidenspointner},
  {Teengarden}, {Atti\'e}, {Cordier}, {Diehl}, {Ferguson}, {Jean}, {von
  Kienlin}, {Paul}, {S\'anchez}, {Schanne}, {Sizun}, {Skinner}, \&
  {Wunderer}}]{sturner03}
{Sturner}, S.~J., {Shrader}, C.~R., {Weidenspointner}, G., {et~al.} 2003,
  \aap\,subm.

\bibitem[{{Trudolyubov}(2001)}]{trudolyubov01}
{Trudolyubov}, S.~P. 2001, \apj, 558, 276

\bibitem[{{Vedrenne} {et~al.}(2003){Vedrenne}, {Roques}, {Sch\"onfelder},
  {Mandrou}, {Lichti}, {von Kielin}, {Cordier}, {Schanne}, {Kn\"odlseder},
  {Skinner}, {Jean}, {Sanchez}, {Caraveo}, {Teegarden}, {von Ballmoos},
  {Bouchet}, {Matteson}, {Boggs}, {Wunderer}, {Leleux}, { Durouchoux}, {Diehl},
  {Strong}, {Cass\'e}, {Clair}, \& {Andr\'e}}]{vedrenne03}
{Vedrenne}, G., {Roques}, J.-P., {Sch\"onfelder}, V., {et~al.} 2003,
  \aap\,subm.

\bibitem[{{Westergaard} {et~al.}(2003){Westergaard}, {Kretschmar}, {Oxborrow},
  {Unknown}, {Smith}, \& {Doe}}]{westergaard03}
{Westergaard}, N.~J., {Kretschmar}, P., {Oxborrow}, C.~A., {et~al.} 2003,
  \aap\,subm.

\bibitem[{{Zdziarski} {et~al.}(2001){Zdziarski}, {Grove}, {Poutanen}, {Rao}, \&
  {Vadawale}}]{zdziarski01}
{Zdziarski}, A.~A., {Grove}, J.~E., {Poutanen}, J., {Rao}, A.~R., \&
  {Vadawale}, S.~V. 2001, \apjl, 554, L45

\end{thebibliography}

\end{document}